\documentclass[conference]{IEEEtran}
\usepackage{graphicx}
\usepackage{xcolor}
\usepackage{textcomp}
\usepackage{cite}
\usepackage{amsmath}
\usepackage{amsfonts}
\usepackage{ascmac}
\usepackage{algorithm}
\usepackage{algorithmicx}
\usepackage{algpseudocode}
\usepackage{subfigure}
\usepackage{amsmath}
\usepackage{amsfonts}
\usepackage{amsmath,amssymb}
\usepackage{bm}
\usepackage{multirow}
\usepackage{mathtools}
\usepackage{textcomp}
\usepackage{url}
\usepackage[utf8]{inputenc}
\usepackage{commath}
\usepackage{tikz}
\usepackage{pgfplots}
\usepackage{pgfplotstable}

\algnewcommand\algorithmicforeach{\textbf{for each}}
\algdef{S}[FOR]{ForEach}[1]{\algorithmicforeach\ #1\ \algorithmicdo}

\newcommand*{\affaddr}[1]{#1} 
\newcommand*{\affmark}[1][*]{\textsuperscript{#1}}

\begin{document}
\title{Unsupervised Recycled FPGA Detection
  \\Using Symmetry Analysis }

\author{
Tanvir Ahmad Tarique\affmark[1], Foisal Ahmed\affmark[2], Maksim Jenihhin\affmark[2], Liakot Ali\affmark[1]\\
\affaddr{\affmark[1]Institute of Information and Communication Technology, Bangladesh University of Engineering
and Technology}\\
Dhaka-1000, Bangladesh, Email: tanvir.tarique5@gmail.com, liakot@iict.buet.ac.bd\\
 \affaddr{\affmark[2]Department of Computer Systems,  Tallinn University of Technology, }\\
19086 Tallinn, Estonia, Email: \{foisal.ahmed, maksim.jenihhin\}@taltech.ee\\
}

\maketitle

\begin{abstract}
  \hspace{1mm}
Recently, recycled field-programmable gate arrays (FPGAs) pose a significant hardware security problem due to the proliferation of the semiconductor supply chain. Ring oscillator (RO) based frequency analyzing technique is one of the popular methods, where most studies used the known fresh FPGAs (KFFs) in machine learning-based detection, which is not a realistic approach. In this paper, we present a novel recycled FPGA detection method by examining the symmetry information of the RO frequency using unsupervised anomaly detection method. Due to the symmetrical array structure of the FPGA, some adjacent logic blocks on an FPGA have comparable RO frequencies, hence our method simply analyzes the RO frequencies of those blocks to determine how similar they are. The proposed approach efficiently categorizes recycled FPGAs by utilizing direct density ratio estimation through outliers detection.
Experiments using Xilinx Artix-7 FPGAs demonstrate that the proposed method accurately classifies recycled FPGAs from 10 fresh FPGAs by x\% fewer computations compared with the conventional method. 
\end{abstract}
\begin{IEEEkeywords}
\small
Recycled FPGA detection, Symmetry analysis, Unsupervised outlier detection, Process variation, Ring oscillator, Direct density ratio estimation
\end{IEEEkeywords}

\section{Introduction}\label{sec:introduction}

Nowadays, recycled field-programmable gate arrays (FPGAs) are a significant counterfeit issue in the IC supply chain due to the increased number of third-party IC vendors. It is reported in study~\cite{R_uzzal} that over 80\% counterfeit electronic components are recycled. Recycled FPGAs have reliability risks and trustworthiness issues owing to aging-induced performance degradation. Meanwhile, presently FPGAs are extensively used in autonomous applications such as UAVs and self-driving cars owing to the excellent performance of AI implementation in safety and critical applications~\cite{foisal_survey}.
If untrusted FPGAs infiltrate mission-critical systems, the system's reliability may suffer, causing significant incidents.

Many methods have been suggested for detecting recycled FPGAs~\cite{R_FPGA,ETS2019_Ahmed,R_FP,foisal_tcad20,alam_tcad19}. The fundamental idea underlying these techniques is to use the ring oscillator (RO) frequency to examine how aging causes circuit characteristics to deteriorate.
The measurement of new FPGA RO frequencies is utilized to train a machine-learning (ML) model. The trained model can identify whether the FPGA under test (FUT) is new or recycled because these frequencies deteriorate with use. These techniques~\cite{R_FPGA,R_FP,foisal_itcasia19,foisal_tcad20} are predicated on the existence of known fresh FPGAs (KFFs). To accurately classify data using ML techniques, FPGA manufacturers must test a high number of KFFs. These methods may not work if there are fewer KFFs available. The work~\cite{alam_tcad19} proposed an unsupervised recycled FPGA detection technique to meet this requirement.

However, because this approach applies the k-means++ clustering algorithm using the measured frequencies as the input vector~\cite{arthur2006k}, its classification accuracy is constrained due to process variation. Furthermore, the logic blocks for RO measurement need to be selected carefully because incorrect and/or inadequate logic blocks reduce classification accuracy. The frequencies of the neighbor block are differentiated exhaustively using the direct density ratio estimation technique to detect the recycled FPGA in~\cite{foisal_iolts20}. However, although this approach classifies the recycled FPGA efficiently, it requires many comparisons within the blocks to find the maximum anomaly score.

\begin{figure}[!t]
\centering
\includegraphics[width=.74\linewidth]{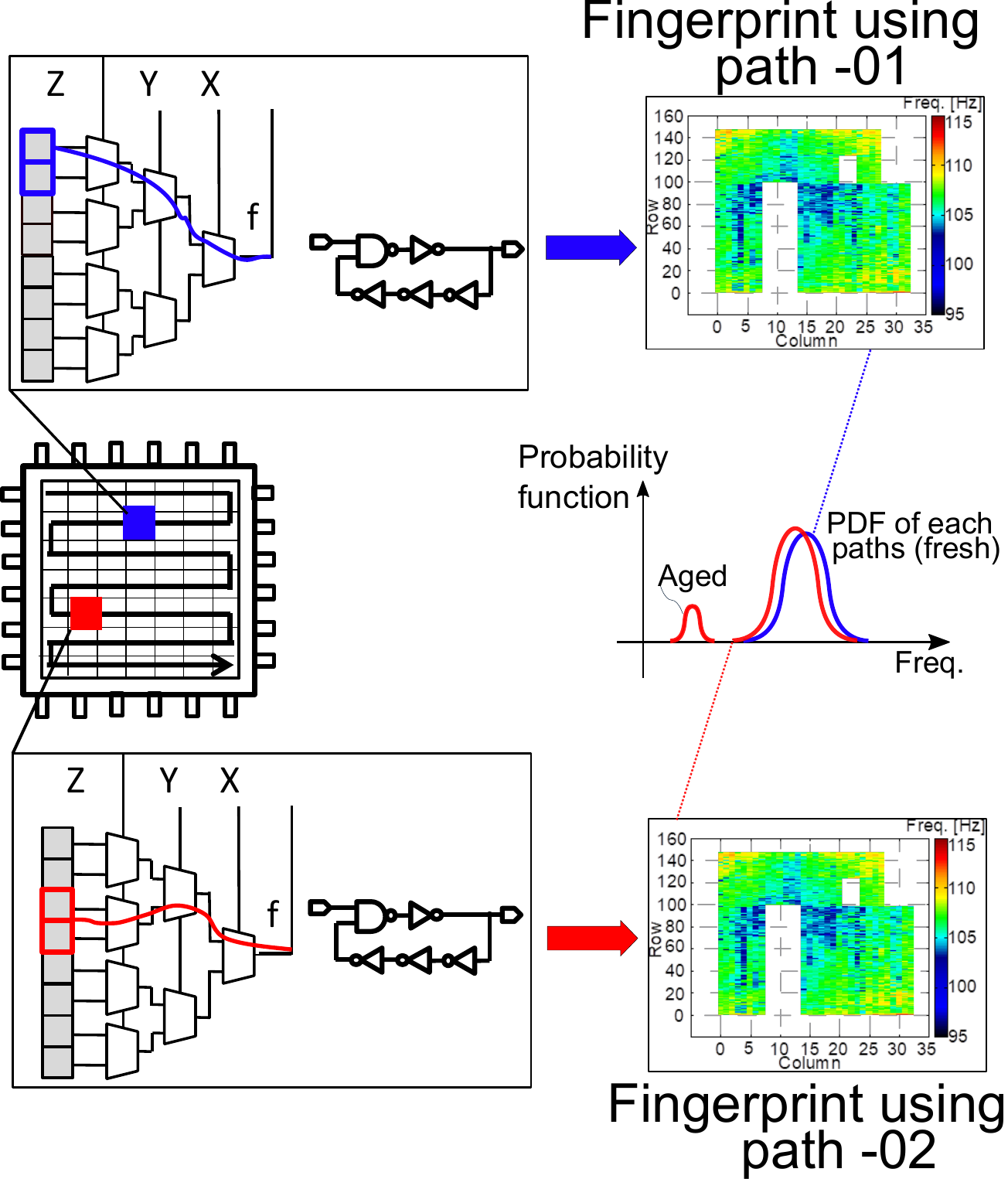}
\caption{Basic idea of the proposed method.}
\label{fig1}
\end{figure}

In this study, we propose a novel method to identify recycled FPGAs without KFF by examining the frequencies on neighboring look-up table (LUT) paths that are symmetrical in structure. Figure~\ref{fig1} depicts the overview concept of the proposed method. During fingerprint (FP) generation  ROs are designed such that each logic configuration block follows the same LUT path as in work~\cite{foisal_tcad20}. The probability density functions (PDFs) of these RO frequencies distribution of some FPs represent similar systematic process variation due to the symmetric structure of the FPGA architecture. Hereafter, we denote these FPs as symmetric path FPs. Due to the delay variations such as aging or malicious insertion, the symmetry relationship can be broken as discussed in~\cite{yoshimizu2014hardware}. We only compare the FPs based on the symmetric path analysis. Therefore, the number of comparisons in the recycled FPGA detection is reduced in this work. While previous work~\cite{foisal_iolts21} exhaustively assessed all comparisons between adjacent columns in the FP, this proposed work only considers comparisons between the symmetric path FP.

The main contribution of this work is outlined as follows:
\begin{itemize}
  \setlength{\itemsep}{-1pt}
\item The proposed recycled FPGA detection analyzes the symmetric relationship  among the fingerprints.
\item On the basis of the symmetric relationship  between the fingerprints, we apply an efficient detection technique employing an unsupervised algorithm by self-referencing using density ratio
estimation.
\item Results from silicon measurements using 10 commercial FPGAs show that the suggested method effectively detects the aged FPGA with $\sim$50\% less computation than the conventional method.
\end{itemize}

The rest of this paper is organized as follows. 
    
In Section~\ref{sec:propose} describes the proposed recycled FPGA
detection method. The experimental procedure and result of the
silicon measurement is discussed in Sec.~\ref{sec:exp}. Finally, we will conclude our paper in Sec.~\ref{sec:conclusion}.

\section{Recycled FPGA Detection using Symmetric Path analysis}\label{sec:propose}

\begin{figure}[!t]
  \centering
  \includegraphics[clip, width=0.7\linewidth]{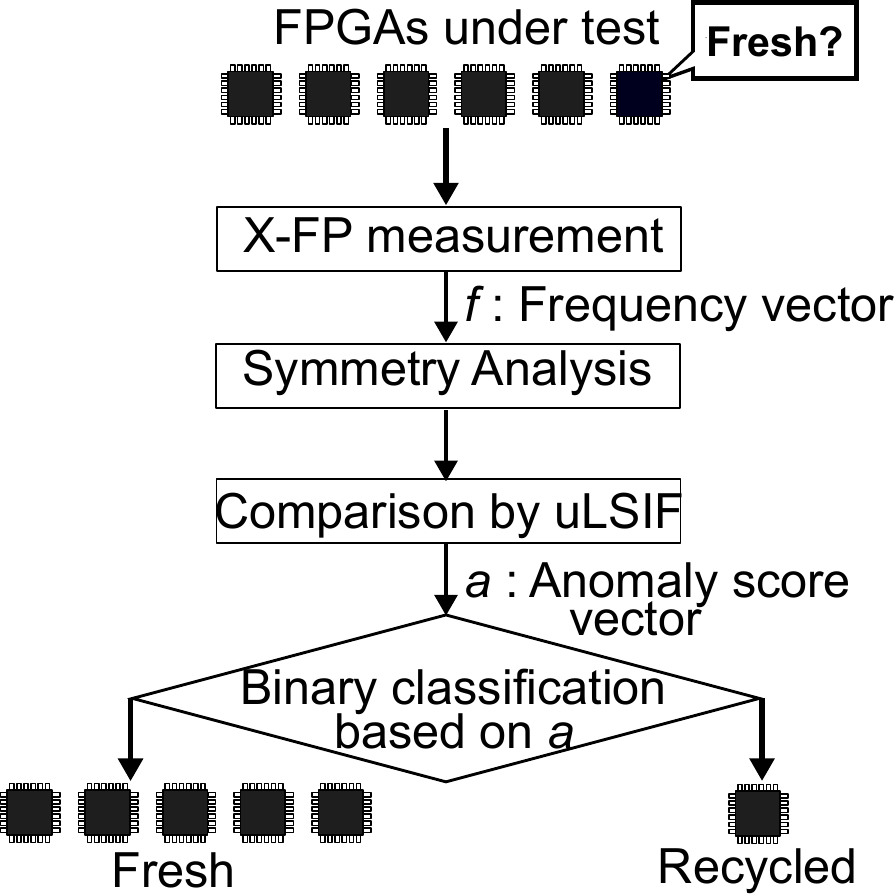}
  \caption{Proposed recycled FPGA detection flow.}
  \label{fig:proposed}
\end{figure}
This unsupervised recycled FPGA detection technique also relies on the
existing standard method~\cite{foisal_iolts21}, and it is outlined
in Fig.~\ref{fig:proposed}. In this work, we intend to reduce the number of computations of the comparison for the anomaly score by the symmetry analysis shown in the figure. The proposed method follows similar steps as~\cite{foisal_iolts21} for the RO measurement and the frequency comparisons in the unsupervised recycled FPGA detection. This method includes an additional step that analyzes the symmetry among the different FPs to get the best match for minimizing the number of comparisons. The anomaly score is derived only from those symmetry FPs. 
For detection, the self-referencing outlier detection technique is designed using unconstrained least-squares importance fitting (uLSIF) algorithm~\cite{hido2011statistical}.

\begin{figure}[!t]
  \centering
  \includegraphics[clip, width=\linewidth]{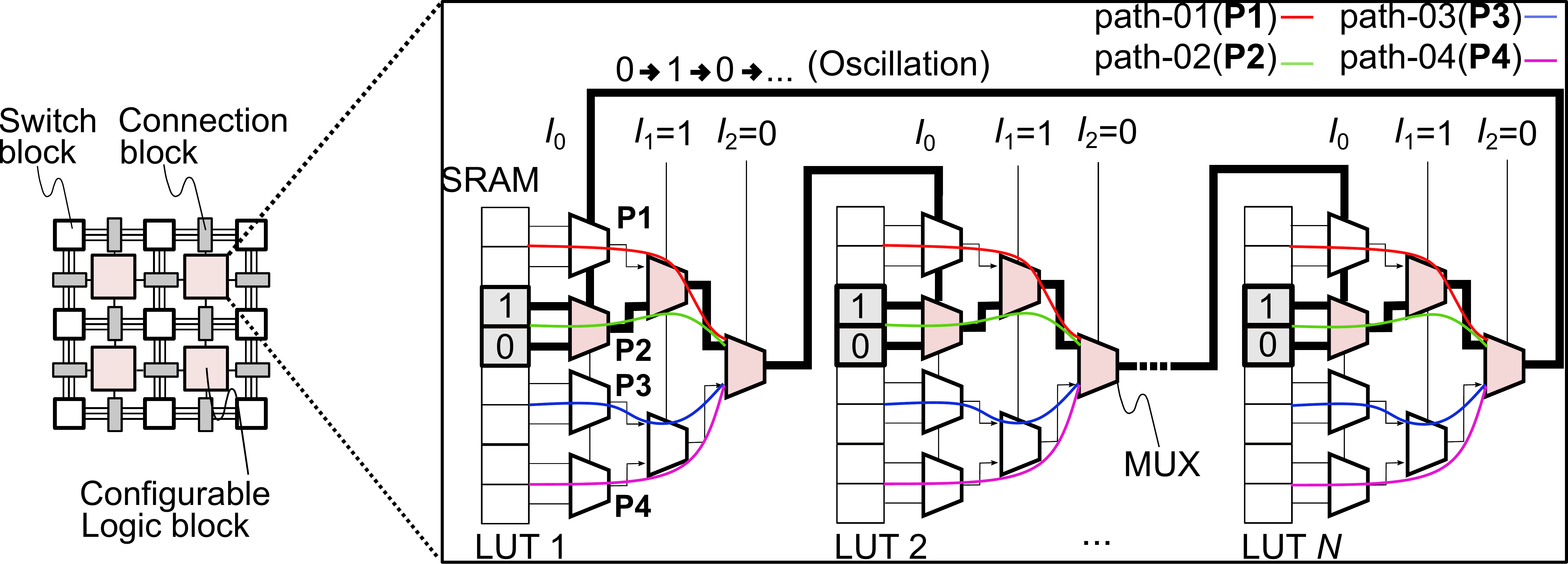}
  \caption{FPGA's basic architecture. Multiple LUTs can be used to construct a multistage RO in a single CLB. The SRAM values are accessible when $I1 = 1$ and $I2 = 0$ are set, causing the highlighted LUT path to oscillate. All LUT paths are characterized when the XP measurement is performed.}
  \label{lut_ro}
\end{figure}

It is easier to understand the RO measurement procedure by looking at the basic FPGA structure shown in Fig.~\ref{lut_ro}. The exhaustive path fingerprinting (X-FP) technique presented in~\cite{foisal_tcad20} is used in this work to design and measure ROs on a FUT. The X-FP measurement method is a combination of two previous works~\cite{R_FP,alam_tcad19}. X-FP characterization, in contrast to exhaustive path (XP) measurement, which requires a process of target CLB selection, executes for all LUT paths of all CLBs in the FPGA, exhaustively capturing the aging result. For instance, if the number of full LUT paths is 4 (P1, P2, P3, and P4), there are four RO measurements that can be possible by accessing four paths (path-01 to path-04) shown in the Figure. Finally, RO measurements are carried out through all CLBs to measure the X-FP that represents all the LUT paths of all the CLBs. If there are $P$ number of paths in a LUT, the total number of X-FP can be represented as $F = F_1, F_2, ..., F_p$, where $F_p$ is the X-FP of the $p$-th path and each FP contains $n$ number of RO measurements.  

For finding the symmetry among the X-FPs, we have utilized a Virtual Probe (VP)-based X-FP estimation by using various sample frequencies as in~\cite{foisal_itcasia19}. The root-mean-square error (RMSE) is calculated between the estimated and measured X-FP. The symmetry path fingerprints are then obtained based on the RMSE values. For instance, the RMSE value of $F_1$ and $F_3$ are very similar, so they are considered symmetry path fingerprints. In this proposed method, the comparisons are performed only on the symmetry path fingerprints. 

This method uses the anomaly score calculated utilizing uLSIF to perform recycled FPGA detection.
For determining the anomaly score, the frequencies of the X-FP measurement are specified as $F_p = f_{p;1}, f_{p;2}, ..., f_{p;n}$, where
$f_{p;n}$ is the RO frequency of the $n$-th RO in the $p$-th path,
and are compared with symmetry path FP based on the RMSE value of the estimated X-FP using the VP technique. The uLSIF algorithm receives the two X-FP vectors of the symmetry path FPs as F and F$^\prime$ in order to determine the anomaly scores. The important contribution of this method is that we do not need to compare all comparisons as in~\cite{foisal_iolts21}. If $C$ is the total number of columns in each X-FP, then the total number of comparisons is required $C\times P/2$ whereas the previous works~\cite{foisal_iolts21} takes total $(C-1)\times P$  comparisons. 
 \section{Experimental Evaluation}\label{sec:exp}


The efficacy  of this technique is assessed by conducting measurements on the Xilinx Artix-7 FPGA~\cite{artix7}, which is manufactured in 28 nm process technology nodes.
 In this experiment, 10 FPGAs (FPGA-01 to FPGA-10) were used.

\subsection{Silicon Measurement}\label{sec:measure}

\begin{figure}[!h]
  \centering
  \includegraphics[clip, width=\linewidth]{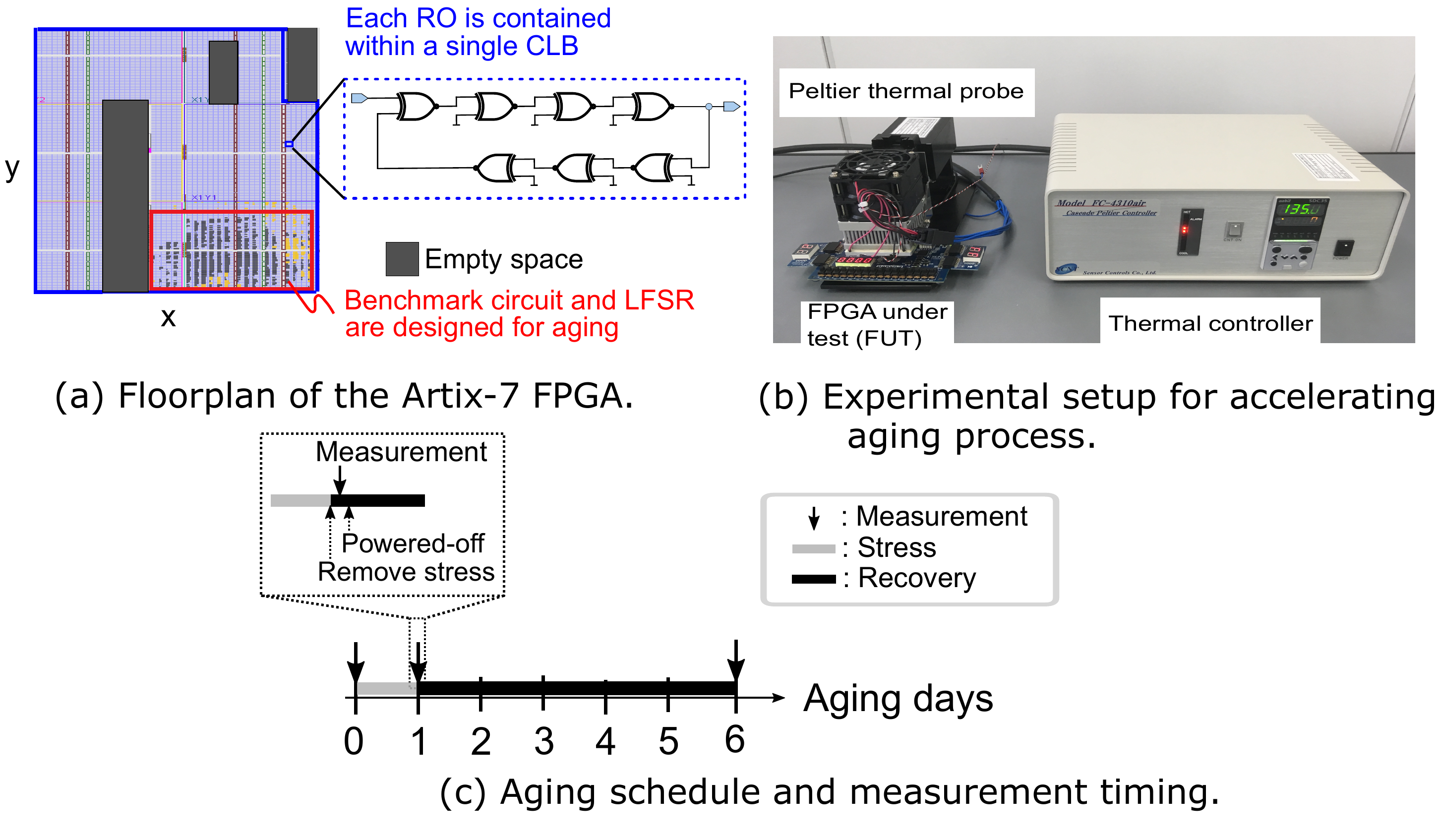}
  \caption{Experimental setup of this work.}
  \label{fig:setup}
\end{figure}


\subsubsection{Measurement Setup}
In this work, we have used a similar experimental setup as in~\cite{foisal_tcad20}.
In the experiment, 7-stage ROs were designed
based on XNOR or XOR logic gates.  We set total of 3,173 ROs in a $33\times120$
    array as shown in Fig.~\ref{fig:setup}(a).  
Using the Xilinx CAD tool Vivado~\cite{vivado}, all ROs are measured, and the results are then combined to create a unique X-FP that illustrates the spatial correlation of the manufacturing process variation.
A total of $32 (= 2^{6-1})$ paths are constructed for XNOR- and XOR-based RO configuration (path-01 to path-32) as Artix-7 FPGA has 6-input LUTs. Thus, to achieve the X-FP measurement total of 32 fingerprint measurements were conducted to cover all 32 paths. We followed the same evaluation system on the 10 FPGAs (FPGA-01 to FPGA-10). Only three FPGAs (FPGA-01 to FPGA-03) were aged from the 10 FPGAs and used as recycled ones.


A Peltier thermal module was utilized in the experimental setup depicted in Fig.~\ref{fig:setup}(b) to accelerate the aging process while the s9234 benchmark circuit from the ISCAS'89 benchmark was being operated~\cite{ISCAS1989_Brglez}. The circuit received random workloads from a 16-bit linear feedback shift register operating at 100\,MHz. In order to take into account the actual scenario in terms of {\it stress} and {\it recovery} state as illustrated in Fig.~\ref{fig:setup}(c),  we used an aging plan for the three FPGAs. The “stress” phases were continued for only 24\,hours (one day). The RO
measurements were carried out only at room temperature after the 5 days recovery phase. The FPGA states at the start and the stop of the recovery phases at the RO frequency measurement are called the “stress
state” and the “recovery state,” respectively. In the ML detection, we have only used 5 days of recovery measurement data to imitate the real-life scenario.

To implement the uLSIF and k-means++ clustering algorithms, we used the Colab Python development environment using Google Cloud.

\subsubsection{Measurement Results}

\begin{figure}[!t]
  \centering
  \includegraphics[clip, width=\linewidth]{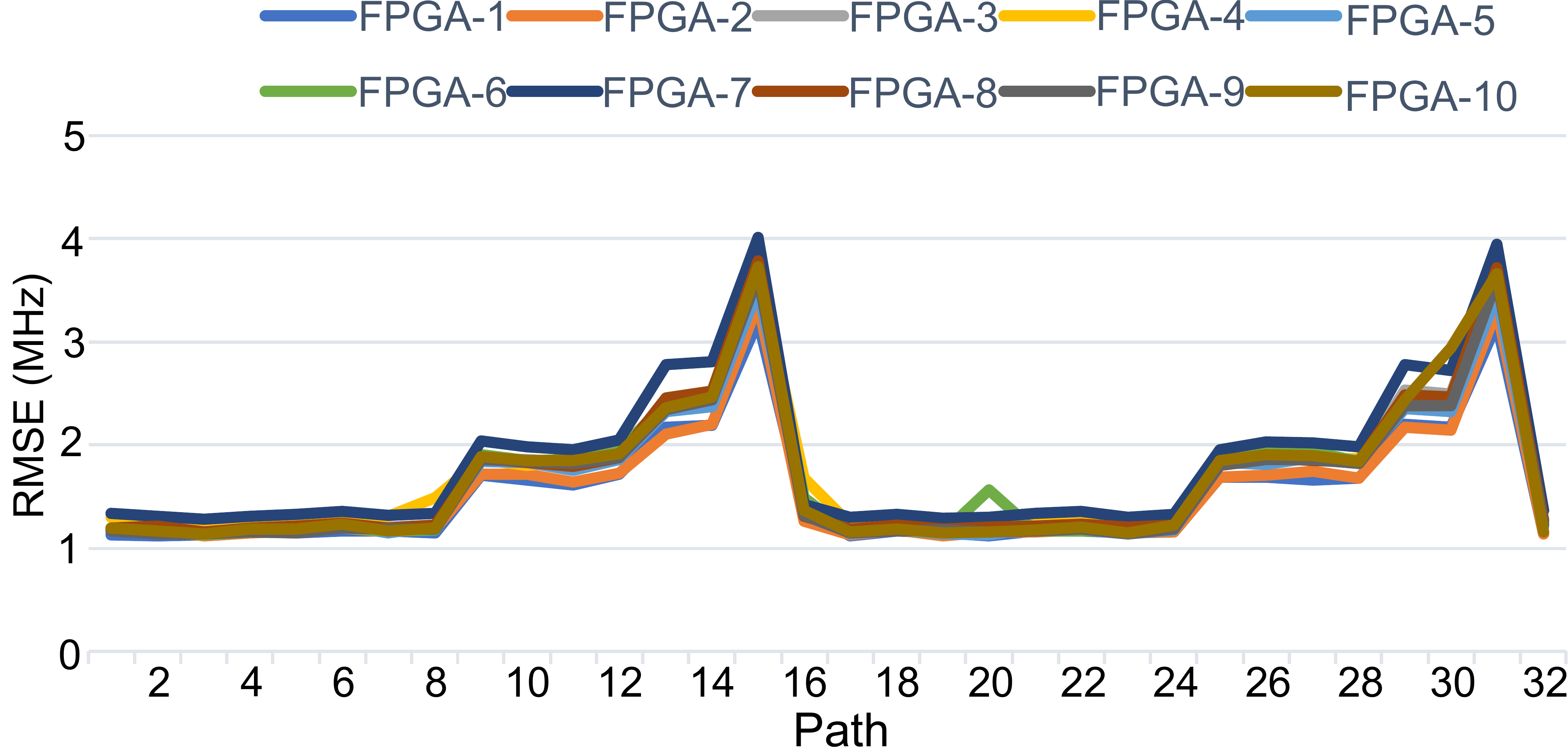}
  \caption{RMSE of 10 fresh FPGAs after applying VP using 10\% samples data.}
  \label{vp_rmse}
\end{figure}

At first, the RMSE data of 10 fresh FPGAs is shown in Fig.~\ref{vp_rmse} at different paths of the X-FP. This RMSE data is obtained from the estimated X-FP after applying the VP technique using 10\% samples over the actual measurement of each X-FP as~\cite{foisal_itcasia19}. From this figure, we can clearly observe the symmetry among various paths. For example, the RMSEs of path-15 and path-31 of FPGA-01 are 3.20 MHz and 3.17 MHz, respectively. As the RMSE values of path-15 and path-31 are very similar, we consider them as symmetry paths and these symmetry paths are used for the comparison (CP) in determining the anomaly score. Thus, based on these RMSE values of 10 fresh FPGAs at 32 paths, the 16 symmetry CPs are considered in this study. The 16 CPs of different symmetry paths are shown in Table~\ref{tab1}.
\begin{table}[!t]
    \centering
         \fontsize{6.5}{6.5}\selectfont
     \caption{\label{tab1}The comparison of different paths for the anomaly score}
    \begin{tabular}{c|c|c|c|c|c|c|c|c}
    \hline
        \textbf{CP no.} & \textbf{CP1} & \textbf{CP2} & \textbf{CP3}& \textbf{CP4}& \textbf{CP5} & \textbf{CP6} & \textbf{CP7}& \textbf{CP8}\\ 
        Paths &1, 17&2, 18 &3, 19&4, 20&5, 21&6, 22&7, 23&8, 24\\ \hline
        \textbf{CP no.} & \textbf{CP9} & \textbf{CP10} & \textbf{CP11} & \textbf{CP12}& \textbf{CP13}& \textbf{CP14} & \textbf{CP15} & \textbf{CP16}\\ 
        Paths &9, 25&10, 26&11, 27&12, 28&13, 29&14, 30&15, 31&16, 32\\ 
     \hline
    \end{tabular}
\end{table}

\begin{figure}[!t]
  \centering
  \includegraphics[clip, width=\linewidth]{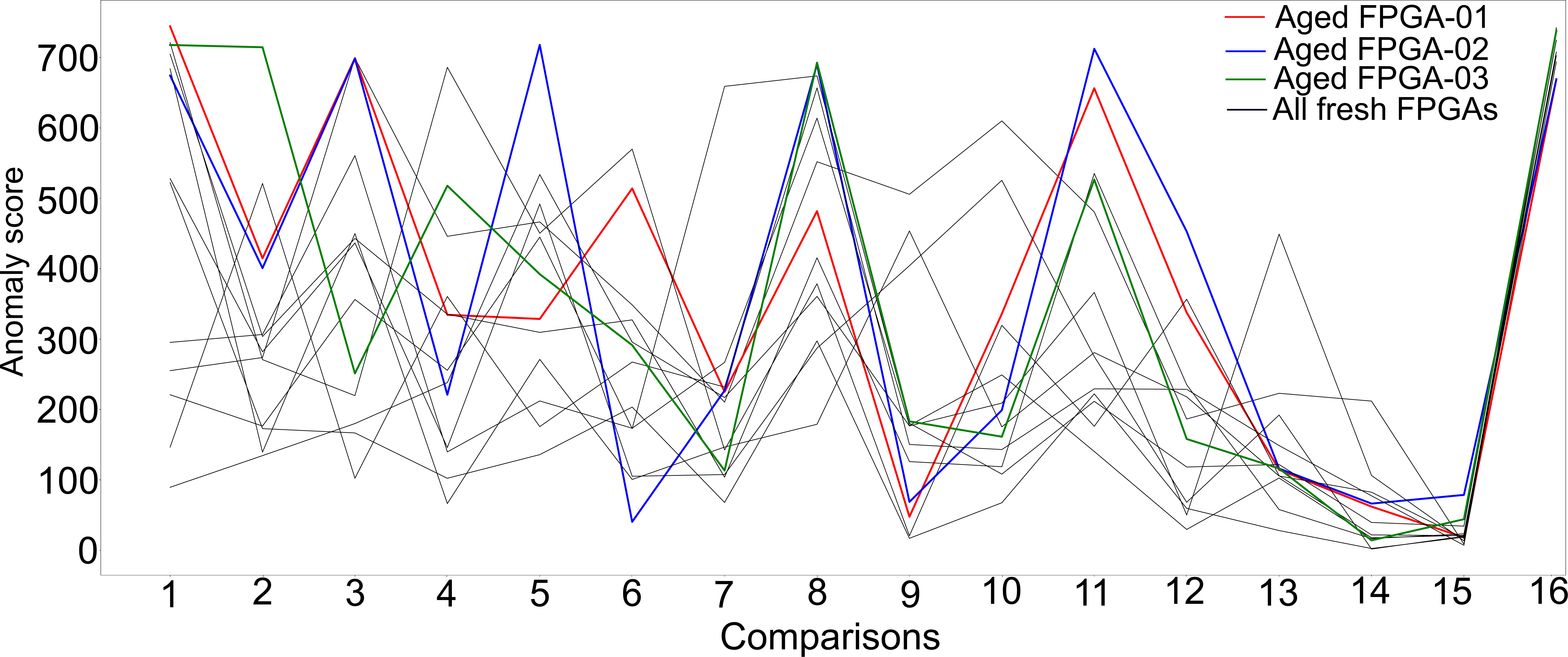}
  \caption{Maximum anomaly scores of the 16 CPs using our proposed approach where used 10 fresh FPGAs and three aged FPGAs.}
  \label{anomaly}
\end{figure}

Based on the CP values shown in Table~\ref{tab1}, we obtained the anomaly score using our proposed method. Figure~\ref{anomaly} presents the anomaly scores of 10 fresh FPGAs (FPGA-01 to FPGA-10) and three aged FPGAs (FPGA-01 to FPGA-03). The vertical axis shows the anomaly scores at 16 different CPs. From this figure, we observe that in most cases the anomaly score of the aged FPGAs is higher than the fresh FPGAs. For instance, aged FPGA-01 (red color) and FPGA-03 (green color) are found high anomaly scores in three CPs (1, 2, and 11) and four CPs (1, 2, 8, and 11), respectively. There are few cases found when fresh FPGAs are shown higher values. As significant anomaly scores are found in the aged FPGAs at different CPs, it is expected that the aged FPGAs could be detected correctly by the unsupervised ML model.

\subsection{Recycled FPGA Detection and Comparison}
\begin{figure}[t]
  \centering
  \includegraphics[clip, width=0.58\linewidth]{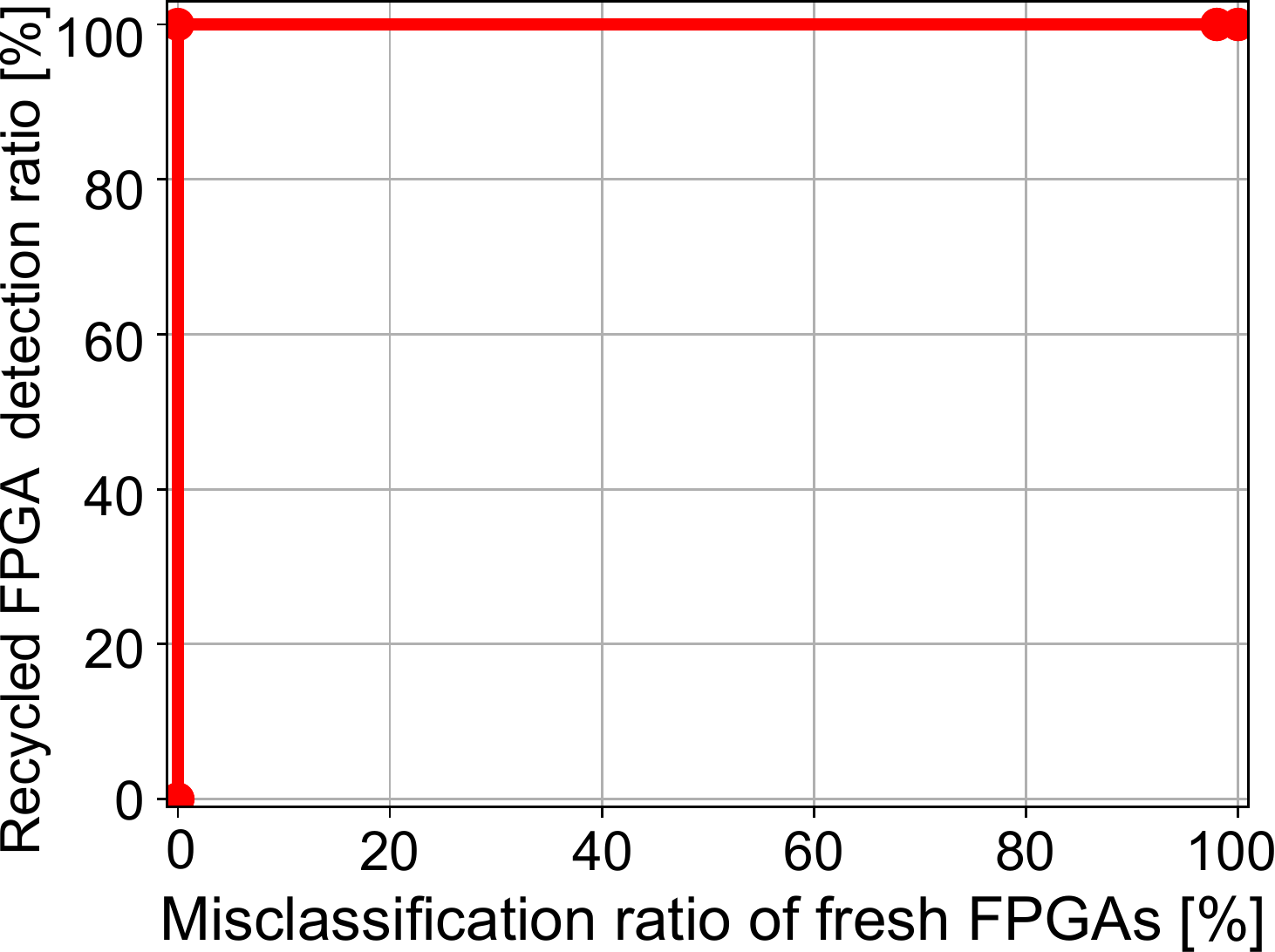}
  \caption{ROC curve of the recycled FPGA detection using 10 fresh FPGAs and 3 aged FPGAs.}
  \label{roc}
\end{figure}
Finally, we have evaluated our proposed recycled FPGA detection using the obtained anomaly score based on the symmetry CPs. The receiver operating characteristics (ROC) curve is shown in Fig.~\ref{roc} to observe the classification results. Here, in the figure, “Recycled FPGA detection ratio” means a true positive rate, and “Misclassification ratio of fresh FPGAs” means a false positive rate. The upper left corner of the ROC curve shows the best performance in the Figure. From the figure, we can observe that in all cases, the proposed technique detects the aged FPGAs accurately. In one case, the fresh FPGA-03 is misclassified due to the effects of process variation. Thus, we can conclude that using our proposed technique, the recycled FPGAs are detected using the unsupervised ML algorithm.

\begin{table}[H]
    \centering
         \fontsize{8}{8}\selectfont
     \caption{\label{tab2}The comparison of the proposed technique}
    \begin{tabular}{c|c|c|c|c}
    \hline
        \textbf{Method} & \textbf{Fresh} & \textbf{Aged} & \textbf{Accuracy}& \textbf{Computations} \\ 
              & \textbf{FPGAs} & \textbf{FPGAs} &  & \textbf{per FPGA}\\ 
        \hline\hline
        Ref~\cite{foisal_iolts21} & 10 & 2 & 100\% & $(C-1)\times P$  \\ 
          & & (6h) &  &   \\\hline
        Proposed & 10 & 3 & 92.31\% & $C\times P/2$ \\
        method &  &(24h) &  &   \\
     \hline
    \end{tabular}
\end{table}

Table~\ref{tab2} shows the comparison with the previous work where our proposed method achieves an almost similar accuracy using around 50\% fewer computations for each FPGA.

\section{Conclusion}\label{sec:conclusion}
In this work, we proposed unsupervised recycled FPGA detection by using symmetry analysis between the X-FPs. The anomaly score is calculated using direct density estimation from the symmetry comparisons. An unsupervised ML algorithm using k-means++ clustering is utilized to detect the recycled FPGAs from fresh samples. The proposed method detects all aged FPGAs by achieving 50\% less computation than the previous work.

\section*{Acknowledgments}
 The authors would like to acknowledge IICT, BUET, Bangladesh for conducting the research using all of its facilities and this work was partly supported by the Estonian Research Council grant PUT PRG1467 “CRASHLESS." We also like to acknowledge the DSL lab at Nara Institute of Science and Technology, Ikoma, Nara, Japan for the use of their experimental data in this research.

\bibliographystyle{IEEEtran}

\end{document}